\documentclass[superscriptaddress, prd, aps,amsmath,amssymb,showpacs,showkeys, onecolumn]{revtex4-2}
\usepackage[dvips]{graphicx,color}
\usepackage{subfig}
\usepackage{times}
\usepackage{xcolor}
\usepackage[%
  colorlinks=true,
  urlcolor=blue,
  linkcolor=red,
  citecolor=blue
]{hyperref}
\usepackage{orcidlink}

\begin{document}
\title{Quasinormal Modes of Black holes in $f(Q)$ gravity}

\author{Dhruba Jyoti Gogoi \orcidlink{0000-0002-4776-8506}}
\email[Email: ]{moloydhruba@yahoo.in}
\affiliation{Department of Physics, Dibrugarh University,
Dibrugarh 786004, Assam, India.}

\author{Ali \"Ovg\"un \orcidlink{0000-0002-9889-342X}}
\email[Email: ]{ali.ovgun@emu.edu.tr}
\affiliation{Physics Department, Eastern Mediterranean University, Famagusta, 99628 
North Cyprus via Mersin 10, Turkey.}

\author{M. Koussour\orcidlink{0000-0002-4188-0572}}
\email[Email: ]{pr.mouhssine@gmail.com}
\affiliation{Quantum Physics and Magnetism Team, LPMC, Faculty of Science Ben
M'sik,\\
Casablanca Hassan II University,
Morocco.}

\begin{abstract}
In this work, we have studied the quasinormal modes of a black hole in a model of the type $f(Q)=\underset{n}{\sum}a_{n}\left(Q-Q_{0}\right)^{n} $ in $f(Q)$ gravity by using a recently introduced method known as Bernstein spectral method and confirmed the validity of the method with the help of well known Pad\'e averaged higher order WKB approximation method. Here we have considered scalar perturbation and electromagnetic perturbation in the black hole spacetime and obtained the corresponding quasinormal modes. We see that for a non-vanishing nonmetricity scalar $Q_0$, quasinormal frequencies in scalar perturbation are greater than those in electromagnetic perturbation scenarios. On the other hand, the damping rate of gravitational waves is higher for electromagnetic perturbation. To confirm the quasinormal mode behaviour, we have also investigated the time domain profiles for both types of perturbations.
\end{abstract}

\keywords{$f(Q)$ Gravity; Gravitational Waves; Quasinormal modes, Black holes}

\maketitle
\section{Introduction}
In recent years, there has been an increase in interest in modified gravity
theories (MGT) in trying to answer various unresolved cosmological puzzles,
such as the accelerated expansion of the universe and the formation of dark
matter. Some of these theories add higher powers of the scalar curvature $R$%
, the Riemann and Ricci tensors, or their derivatives to General Relativity
(GR). Some of these efforts include the $f(R)$ and $f(G)$ theories, where $R$
and $G$ are the Ricci scalar and the Gauss--Bonnet topological invariant,
respectively \cite{Starobinsky1, fG}. Certain viable $f(R)$ gravity models
have been proposed \cite{fR1}, which support the unification of early-time
inflation and late-time acceleration. Viable $f(R)$ gravity models can also
be used to solve dark matter issues \cite{fR2, fR3, fR4}. In this paper, we
examine the quasinormal modes of black holes in the recently suggested $f(Q)$
theory of gravity, where $Q$ is the non-metricity scalar \cite{Q0, Q1}. The
non-metricity $Q$ of the metric mathematically describes the variation in
the length of a vector in a parallel transport process, and it is the key
geometric variable describing the properties of the gravitational effect.

Generally, the gravitational effects in the space-time manifold can be
described using three geometrical objects: curvature $R$, torsion $T$, and
non-metricity $Q$. In GR, gravitational effects are assigned to space-time
curvature. Another two scenarios, torsion, and non-metricity offer the
equivalent representation of GR, and the associated gravity is known as the
teleparallel and symmetric teleparallel equivalent of GR. The $f(R)$ theory
is a curvature-based extension of GR with zero torsion and non-metricity.
Likewise, the $f(T)$ theory with zero non-metricity and curvature is an
extension of torsion-based gravity (the teleparallel equivalent of GR) \cite%
{T1, T2, T3,Bahamonde:2021gfp,Adak:2005cd,Adak:2004uh,Adak:2008gd}. Finally, the $f(Q)$ gravity theory generalizes GR's symmetric
teleparallel (ST) equivalent with zero torsion and curvature. The expanding
history of the Universe in $f(Q)$ gravity attracts instant attention as one
of the key drivers for this extension \cite{Kou1, Kou2, Kou3, Kou4}. Many
observational data sets have been used to test the $f(Q)$ gravity in recent
years, as seen in the study of Lazkoz et al. \cite{Q2}. The authors used
data from the expansion rate, Type Ia Supernovae (SNe Ia), Quasars,
Gamma-Ray Bursts, Baryon Acoustic Oscillations (BAO), and the Cosmic
Microwave Background (CMB) to constrain $f(Q)$ gravity. Mandal et al. \cite%
{Q3} also shown the validity of $f(Q)$ cosmological models in terms of
energy conditions. The authors of such a study established the so-called
embedding approach, which allows non-trivial contributions from the
non-metricity function to be included in the energy conditions. Furthermore,
we can witness an increasing interest in $f(Q)$ gravity in the study of
astrophysical objects. Wang et al. \cite{Q6} also studied the static and
spherically symmetric solutions for $f(Q)$ gravity in an anisotropic fluid. In 
\cite{Q4}, black holes in $f(Q)$ gravity have been studied. The authors of 
\cite{Q5} investigated the usage of spherically symmetric configurations in $%
f(Q)$ gravity. Hassan et al. \cite{Q7} used linear equation of state (EoS)
and anisotropic relations to examine wormhole geometries in $f(Q)$ gravity.
They discovered exact solutions to the linear model and confirmed the
presence of a modest quantity of exotic matter necessary for a traversable
wormhole through VIQ. Mustafa et al. have also derived wormhole solutions
from the Karmarkar condition and demonstrated the possibility of creating
traversable wormholes while maintaining the energy conditions in \cite{Q8}.
Furthermore, Banerjee et al.\cite{Q9} examined the behaviour of energy
conditions under constant redshift function by assuming some particular
shape functions and confirmed that wormhole solutions could not exist for $%
f(Q)=Q+\alpha Q^{2}$ model. Ref. \cite{Q10} recently examined a class of static
spherically symmetric solutions in $f(Q)$ gravity. Traversable wormholes with charge and non-commutative geometry are studied in Ref.  \cite{Q11}.

Black holes are the cleanest objects in the Universe and are highly 
associated with the generation of gravitational waves. Quasinormal modes are 
a fascinating and important aspect of black hole physics. They are 
oscillations of a black hole that are damped over time and characterized by 
complex frequencies. The term ``quasinormal" refers to the fact that these 
modes are not exactly normal modes, which would oscillate indefinitely. 
Instead, they die out due to the presence of dissipative effects, such as the 
emission of gravitational waves. Quasinormal modes are basically complex 
values that correspond to the emission of gravitational waves from compact 
and massive objects in the Universe 
\cite{Vishveshwara, Press, Chandrasekhar_qnms}. The real component of the 
quasinormal modes signifies the emission frequency, while the imaginary 
component pertains to its decay. Quasinormal modes are significant because 
they encode information about the black hole's properties, such as its mass, 
angular momentum, and the properties of the surrounding spacetime. 
Additionally, the study of quasinormal modes provides insight into the nature 
of black holes and the strong gravity regime, which is difficult to probe 
using other means. These modes are important for understanding the structure 
and evolution of black holes, as well as their role in astrophysical 
phenomena such as gravitational wave signals.

The properties of gravitational waves and quasinormal modes of black holes 
have been extensively explored in various modified gravity theories in recent 
years \cite{Ma, gogoi1, gogoi2, Liang_2017, qnm_bumblebee, gogoi3, Graca, Zhang2, lopez2020, Liang2018, Hu, hemawati2022, gogoi4, Ovgun:2017dvs,Rincon:2018sgd,Panotopoulos:2019qjk,Panotopoulos:2020mii, Rincon:2021gwd, Gonzalez:2021vwp,Daghigh:2011ty, Daghigh:2008jz, Zhidenko:2003wq, Zhidenko:2005mv, Konoplya:2011qq, Hatsuda:2019eoj, Eniceicu:2019npi,Lepe:2004kv, Chabab:2017knz, Chabab:2016cem,Okyay2022, Jusufi18}. In a recent study, quasinormal modes, black hole 
shadow, greybody bounds etc., are studied extensively in dyonic modified 
Maxwell black holes \cite{Pantig:2022gih}. In another study, the quasinormal 
modes have been investigated for the Kerr-like black bounce spacetime under 
the scalar field perturbations \cite{Yang:2022xxh}. Another research analyses 
the quasinormal modes of hairy black holes produced by gravitational 
decoupling for massless scalar fields, electromagnetic fields, and 
gravitational perturbations \cite{Yang:2022ifo}. In this work, the equation 
for the effective potential of these three perturbations is determined within 
the spacetime of the hairy black holes. The time evolution for the three 
perturbations is also studied, and the quasinormal mode frequencies are 
calculated using the Prony method based on the time-domain profiles. In a 
recent publication, the authors demonstrate how quasinormal modes are 
produced by the perturbations of massive scalar fields in a curved background 
through the use of artificial neural networks \cite{Ovgun:2019yor}. They 
design a specific algorithm for the feed-forward neural network method to 
calculate the quasinormal modes that meet specific boundary conditions. To 
verify the accuracy of the method, they examine two black hole spacetimes 
whose quasinormal modes are well-established: the 4D pure de Sitter (dS) and 
five-dimensional Schwarzschild anti-de Sitter (AdS) black holes. Apart from 
black holes, quasinormal modes are also studied for the wormholes in 
different frameworks \cite{Gonzalez:2022ote, gogoi_wormhole}. In this study, 
we shall investigate the quasinormal modes of black holes in $f(Q)$ gravity. 
One may note that the studies of black hole solutions in this theory of 
gravity are not very old. Until now, very few papers have dealt with the 
black holes in the framework of $f(Q)$ gravity \cite{Q6,Q12}. Quasinormal 
modes, as well as thermodynamics of black holes in $f(Q)$ gravity, are still 
unexplored. Therefore, being motivated by the previous studies, in this work, 
we shall investigate the quasinormal modes and time domain profiles of scalar 
and electromagnetic perturbations of a static black hole in the framework of 
$f(Q)$ gravity theory.
There are two main prime objectives of this study. The first one is 
investigating the behaviour of quasinormal modes in a recently obtained black 
hole solution in $f(Q)$ gravity. This study will enable us to see whether the 
presence of nonmetricity can be probed via quasinormal modes. Moreover, the 
impacts of the black hole parameters on the quasinormal spectrum will also be 
investigated thoroughly. The second one is verifying a newly introduced 
method to calculate the quasinormal modes, which is known as the Bernstein 
spectral method. To verify the method in this study, we shall use a well-
known method, the 6th-order Pad\'e averaged WKB approximation method.

The paper is organized as follows: In Sec. \ref{sec2}, we briefly review the 
$f(Q)$ gravity. Sec. \ref{sec3} is devoted to studying the field equations in 
$f(Q)$ gravity and possible vacuum black hole solutions. In Sec. \ref{sec4}, 
we briefly investigate the properties of the black hole solution with the 
help of associated scalars. Sec. \ref{sec5} deals with the scalar and 
electromagnetic perturbation and associated quasinormal modes. In Sec. 
\ref{sec6}, the time evolution profiles of the perturbations are 
investigated. Finally, we conclude the paper with a brief discussion of the 
results and future prospects in Sec. \ref{sec7}.

Throughout the whole paper, we have used $G=c=1$.

\section{A brief overview of $f(Q)$ gravity}

\label{sec2}

Weyl geometry is a significant extension of Riemannian geometry, which
serves as the mathematical foundation for GR. According to Weyl geometry,
during a parallel transport over a closed path, an arbitrary vector will not
only change the direction but also the length. As a result, in Weyl's
theory, the covariant derivative~of the metric tensor is non-zero, and this
feature can be represented mathematically in terms of a new geometric
quantity named non-metricity $Q$. Thus, the non-metricity tensor $Q_{\gamma
\mu \nu }$ can be defined as the covariant derivative of the metric tensor $%
g_{\mu \nu }$ with regard to the general affine connection $\overline{\Gamma 
}_{~\mu \gamma }^{\sigma }$, and it can be expressed as \cite{Q0, Q1, Q13},%
\begin{equation}
Q_{\gamma \mu \nu }=-\nabla _{\gamma }g_{\mu \nu }=-\frac{\partial g_{\mu
\nu }}{\partial x^{\gamma }}+g_{\nu \sigma }\overline{\Gamma }_{~\mu \gamma
}^{\sigma }+g_{\sigma \mu }\overline{\Gamma }_{~\nu \gamma }^{\sigma }.
\label{2f}
\end{equation}

In this situation, the general affine connection is represented by a Weyl
connection and is divided into two independent components as shown below,%
\begin{equation}
\overline{\Gamma }_{\ \mu \nu }^{\gamma }=\Gamma _{\ \mu \nu }^{\gamma
}+L_{\ \mu \nu }^{\gamma },
\end{equation}%
where the first term is the usual Levi-Civita connection of the metric $%
g_{\mu \nu }$, as defined by the standard formulation,%
\begin{equation}
\Gamma _{~\mu \nu }^{\gamma }\equiv\frac{1}{2}g^{\gamma \sigma }\left( \frac{%
\partial g_{\sigma \nu }}{\partial x^{\mu }}+\frac{\partial g_{\sigma \mu }}{%
\partial x^{\nu }}-\frac{\partial g_{\mu \nu }}{\partial x^{\sigma }}\right)
.
\end{equation}

The second term, which represents the disformation tensor due to the
non-metricity of space-time is written as, 
\begin{equation}
L_{~\mu \nu }^{\gamma }\equiv\frac{1}{2}g^{\gamma \sigma }\left( Q_{\nu \mu
\sigma }+Q_{\mu \nu \sigma }-Q_{\gamma \mu \nu }\right) =L_{~\nu \mu
}^{\gamma }.
\end{equation}%
Furthermore, as a function of the disformation tensor, the contraction of
the non-metricity tensor gives the non-metricity scalar,%
\begin{equation}
Q\equiv -g^{\mu \nu }\left( L_{\ \ \beta \mu }^{\alpha }L_{\ \ \nu \alpha
}^{\beta }-L_{\ \ \beta \alpha }^{\alpha }L_{\ \ \mu \nu }^{\beta }\right) .
\end{equation}

The Weyl geometry can be extended by accounting for space-time torsion,
yielding the Weyl-Cartan spaces with torsion. The general affine connection
in the Weyl-Cartan geometry can be divided into three independent components
as,%
\begin{equation}
\overline{\Gamma }_{\ \mu \nu }^{\gamma }=\Gamma _{\ \mu \nu }^{\gamma
}+L_{\ \mu \nu }^{\gamma }+K_{\ \mu \nu }^{\gamma },
\end{equation}%
where the third term is contortion, described in terms of the torsion tensor 
$T_{\ \mu \nu }^{\gamma }\equiv 2\overline{\Gamma }_{~[\mu \nu ]}^{\gamma }$%
\ as,%
\begin{equation}
K_{\ \mu \nu }^{\gamma }\equiv \frac{1}{2}g^{\gamma \sigma }\left( T_{\mu
\sigma \nu }+T_{\nu \sigma \mu }+T_{\sigma \mu \nu }\right) .
\end{equation}

Further, the relation between curvatures tensor $R_{\sigma \mu \nu }^{\rho }$
and $\mathring{R}_{\sigma \mu \nu }^{\rho }$ corresponding to the connection 
$\overline{\Gamma }_{\ \mu \nu }^{\gamma }$ and $\Gamma _{\ \mu \nu
}^{\gamma }$ is,%
\begin{equation}
R_{\sigma \mu \nu }^{\rho }=\mathring{R}_{\sigma \mu \nu }^{\rho }+\mathring{%
\nabla}_{\mu }L_{\nu \sigma }^{\rho }-\mathring{\nabla}_{\nu }L_{\mu \sigma
}^{\rho }+L_{\mu \lambda }^{\rho }L_{\nu \sigma }^{\lambda }-L_{\nu \lambda
}^{\rho }L_{\mu \sigma }^{\lambda },
\end{equation}%
\begin{equation}
R_{\sigma \nu }=\mathring{R}_{\sigma \nu }+\frac{1}{2}\mathring{\nabla}_{\nu
}Q_{\sigma }+\mathring{\nabla}_{\rho }L_{\nu \sigma }^{\rho }-\frac{1}{2}%
Q_{\lambda }L_{\nu \sigma }^{\lambda }-L_{\sigma \lambda }^{\rho }L_{\rho
\sigma }^{\lambda },  \label{2g}
\end{equation}%
and the scalar curvature relation,%
\begin{equation}
R=\mathring{R}+\mathring{\nabla}_{\lambda }Q^{\lambda }-\mathring{\nabla}%
_{\lambda }\tilde{Q}^{\lambda }-\frac{1}{4}Q_{\lambda }Q^{\lambda }+\frac{1}{%
2}Q_{\lambda }\tilde{Q}^{\lambda }-L_{\rho \nu \lambda }L^{\lambda \rho \nu
},  \label{2h}
\end{equation}%
where $\mathring{\nabla}$ is the covariant derivative operator associated
with the Levi-Civita connection $\Gamma _{\ \mu \nu }^{\gamma }$. In Symmetric Teleparallel Gravity (STG) 
case, the curvature-free, and torsion-free requirements restrict the affine
general connection. The curvature-free condition necessitates that the
Riemann tensor $R_{\sigma \mu \nu }^{\rho }\left( \overline{\Gamma }\right) $
be zero. So because the Riemann tensor disappears, the parallel transport
represented by the covariant derivative $\nabla $ and its corresponding
affine connection $\overline{\Gamma }_{\ \mu \nu }^{\gamma }$ is path
independent. In addition to the requirement of zero curvature, this theory
imposes a torsionless restriction on the connection, i.e. $T_{\ \mu \nu
}^{\gamma }=0$, such that gravitation is completely ascribed to
non-metricity in STG. The general affine connection is symmetric in its
lower indices because the torsion tensor disappears.

As explained previously, in order to obtain the STG, two constraints must be
added to the generic affine connection: $R_{\sigma \mu \nu }^{\rho }\left( 
\overline{\Gamma }\right) =0$ and $T_{\ \mu \nu }^{\gamma }=0$. These
constraints permit the choice of a coordinate system $\left\{ y^{\mu
}\right\} $, in which the affine connection $\overline{\Gamma }_{\ \mu \nu
}^{\gamma }\left( y^{\mu }\right) $ disappears i.e. $L_{\ \mu \nu }^{\gamma
}=-\Gamma _{\ \mu \nu }^{\gamma }$, resulting in the so-called coincident
gauge \cite{Q0}. Thus, in any other coordinate system $\left\{ x^{\mu }\right\} $, the
affine connection takes the form,%
\begin{equation}
\overline{\Gamma }_{\ \mu \nu }^{\gamma }\left( x^{\mu }\right) =\frac{%
\partial x^{\gamma }}{\partial y^{\beta }}\partial _{\mu }\partial _{\nu
}y^{\beta }.
\end{equation}

Because there exists a coordinate system $\left\{ y^{\mu }\right\} $ in
which the affine connection disappears, we may always imagine that we are
working in this special coordinate system with metric as the only
fundamental variable. In this case, the covariant derivative $\nabla
_{\gamma }$\ reduces to the partial derivative $\partial _{\gamma }$. As a
result, in the coincident gauge coordinate, we obtain 
\begin{equation}
Q_{\gamma \mu \nu }=-\partial _{\gamma }g_{\mu \nu },  \label{2i}
\end{equation}%
while in another coordinate system, 
\begin{equation}
Q_{\gamma \mu \nu }=-\partial _{\gamma }g_{\mu \nu }-2\overline{\Gamma }%
_{\alpha (\mu }^{\gamma }g_{\nu )\lambda }.  \label{2j}
\end{equation}

Finally, the non-metricity scalar can be used to express the action for the $%
f(Q)$ gravity as \cite{Q0},%
\begin{equation}
S=\int {d^{4}x\tau }\left( \frac{1}{2}{f(Q)+L_{m}}\right) .  \label{2k}
\end{equation}

Here, $L_{m}$ denotes the Lagrangian density of matter and $\tau $ denotes $%
\tau =det(g_{\mu \nu })=\sqrt{-g}$. Similarly to $f(R)$ gravity, ${f(Q)}$
will be responsible for the deviation from GR, where, for example, if the
function $f\left( Q\right) $ is considered to be $-Q$, we recover the
so-called Symmetric Teleparallel Equivalent to GR (STEGR). Since the symmetry of the metric tensor $g_{\mu \nu }$, we
can just derive two independent traces from the non-metricity tensor $%
Q_{\gamma \mu \nu }$,%
\begin{equation}
Q_{\gamma }\equiv Q_{\gamma \ \ \ \mu }^{\ \ \mu },\ \ \ \ \tilde{Q}_{\gamma
}\equiv Q_{\ \ \gamma \mu }^{\mu }.  \label{2l}
\end{equation}

Also, it will be convenient to introduce the non-metricity conjugate defined
as, 
\begin{equation}
\hspace{-0.5cm}P_{\ \ \mu \nu }^{\gamma }\equiv \frac{1}{4}\bigg[-Q_{\ \ \mu
\nu }^{\gamma }+2Q_{\left( \mu \ \ \ \nu \right) }^{\ \ \ \gamma }+Q^{\gamma
}g_{\mu \nu }-\widetilde{Q}^{\gamma }g_{\mu \nu }-\delta _{\ \ (\mu
}^{\gamma }Q_{\nu )}\bigg]=-\frac{1}{2}L_{\ \ \mu \nu }^{\gamma }+\frac{1}{4}%
\left( Q^{\gamma }-\widetilde{Q}^{\gamma }\right) g_{\mu \nu }-\frac{1}{4}%
\delta _{\ \ (\mu }^{\gamma }Q_{\nu )}.
\end{equation}

The scalar of non-metricity is calculated as follows: 
\begin{equation}
Q=-Q_{\gamma \mu \nu }P^{\gamma \mu \nu }=-\frac{1}{4}\big(-Q^{\gamma \nu
\rho }Q_{\gamma \nu \rho }+2Q^{\gamma \nu \rho }Q_{\rho \gamma \nu
}-2Q^{\rho }\tilde{Q}_{\rho }+Q^{\rho }Q_{\rho }\big).
\end{equation}

Further, the stress-energy momentum tensor for cosmic matter content is
determined by%
\begin{equation}
T_{\mu \nu }\equiv -\frac{2}{{\tau }}\frac{\delta ({\tau }L_{m})}{\delta
g^{\mu \nu }}.  \label{2o}
\end{equation}

The gravitational field equations derived by varying action (\ref{2k}) with
regard to the metric $g_{\mu \nu }$ are shown below, 
\begin{widetext}
\begin{equation}\label{2p}
\frac{2}{\sqrt{-g}}\nabla_\gamma (\sqrt{-g}f_Q P^\gamma\:_{\mu\nu}) + \frac{1}{2}g_{\mu\nu}f+f_Q(P_{\mu\gamma\beta}Q_\nu\:^{\gamma\beta} - 2Q_{\gamma\beta\mu}P^{\gamma\beta}\:_\nu) = -T_{\mu\nu}.
\end{equation}
\end{widetext}

For the purpose of simplicity, we designate $f_{Q}=\frac{df}{dQ}$.

Again, in the absence of hypermomentum \cite{Q13}, we get the connection field
equations by varying the gravitational action (\ref{2k}) with regard to the
connection, 
\begin{equation}
\nabla _{\mu }\nabla _{\nu }(\sqrt{-g}f_{Q}P^{\mu \nu }{}_{\gamma })=0.
\label{2q}
\end{equation}

\section{Field Equations and a special case of vacuum black hole solution}

\label{sec3}

In this study, we shall consider the following static and spherically
symmetric ansatz to obtain a static black hole solution. 
\begin{equation}
ds^2 = - e^{a(r)} dt^2 + e^{b(r)} dr^2 + r^2 d \Omega^2,
\end{equation}
where the term $d \Omega^2 = d \theta^2 + \sin^2 \theta d \phi^2$. For this
ansatz, the nonmetricity scalar $Q$ can be written as \cite{Q6} 
\begin{equation}  \label{Q}
Q(r) = - \frac{2 e^{-b(r)}}{r} \left( a^{\prime -1} \right).
\end{equation}
Following Ref. \cite{Q6}, we start with a constant nonmetricity scalar $Q= Q_0$%
, for which \eqref{Q} can be written as 
\begin{equation}  \label{aprime}
a^{\prime }(r) = - \frac{Q_0 r}{2} e^{b(r)} - r^{-1}.
\end{equation}
Now, the components of the field equation can be written as 
\begin{align}  \label{Q0-pt}
\rho&=\frac{f\left(Q_{0}\right)}{2}-f_{Q}\left(Q_{0}\right)\left[Q_{0}+\frac{%
1}{r^{2}}+\frac{e^{-b}}{r}\left(a^{\prime}+b^{\prime}\right)\right] \,, \\
p_{r} &=- \frac{f(Q_{0})}{2} + f_Q(Q_{0})\left(Q_{0}+\frac{1}{r^2}\right) \,
, \\
p_{t} &= - \frac{f(Q_{0})}{2} + f_Q(Q_{0})\left\{ \frac{Q_{0}}{2}-e^{-b}%
\left[\frac{a^{\prime\prime}}{2}+\left(\frac{a^{\prime}}{4}+\frac{1}{2r}%
\right) (a^{\prime}-b^{\prime})\right]\right\} \, .
\end{align}

Now, we shall consider the vacuum case with $\rho = p_r = p_t = 0$, in the
above field equations which reduces them to 
\begin{align}
    \label{vac01}
    0&=f_{Q}\left(Q_{0}\right)\frac{e^{-b}}{r}\left(a^{\prime}+b^{\prime}\right)
    \,,\\
    \label{vac02}
    0&=-\frac{f\left(Q_{0}\right)}{2}+f_{Q}\left(Q_{0}\right)\left(Q_{0}+\frac{1}{r^{2}}\right)
    \,,\\
    \label{vac03}
    0&=f_{Q}\left(Q_{0}\right)\left\{ \frac{Q_{0}}{2}+\frac{1}{r^{2}}+e^{-b}\left[\frac{a^{\prime\prime}}{2}+\left(\frac{a^{\prime}}{4}+\frac{1}{2r}\right)\left(a^{\prime}-b^{\prime}\right)\right]\right\}
    \,.
\end{align}
From the Eq.~\eqref{vac02}, it is possible to have 
\begin{align}  \label{condition01}
f_{Q}\left(Q_{0}\right) = 0 \, , \ f(Q_{0}) = 0 \, .
\end{align}
These conditions suggest the following form of $f(Q)$ as given by, 
\begin{align}  \label{fQmodel}
f(Q)=\underset{n}{\sum}a_{n}\left(Q-Q_{0}\right)^{n} \, ,
\end{align}
where $a_{n}$ represents arbitrary model parameters. Hence, we have seen
that for $f(Q)$ gravity to have nontrivial spacetime solutions, the model
should satisfy the conditions \eqref{condition01}.

Now, we fix $e^{a(r)}=1-\frac{2M}{r}$, which is the normal Schwarzschild
black hole metric function with black hole mass $M$. With this assumption,
Eq.~\eqref{aprime} results 
\begin{align}
e^{b}=-\frac{2}{Q_{0}r\left(r-2M\right)} \,,
\end{align}
where one must have $Q_{0}<0$, and the black hole metric is found to be \cite{Q6}
\begin{align}  \label{Q0-vac-cas-metric}
ds^{2}=-\left(1-\frac{2M}{r}\right)dt^{2}+\frac{-2}{Q_{0}r^{2}}\left(1-\frac{%
2M}{r}\right)^{-1}dr^{2}+r^{2}d\Omega^{2} \,.
\end{align}

\section{Properties of the black hole solution}

\label{sec4} 
\begin{figure}[!h]
\centerline{
   \includegraphics[scale = 0.45]{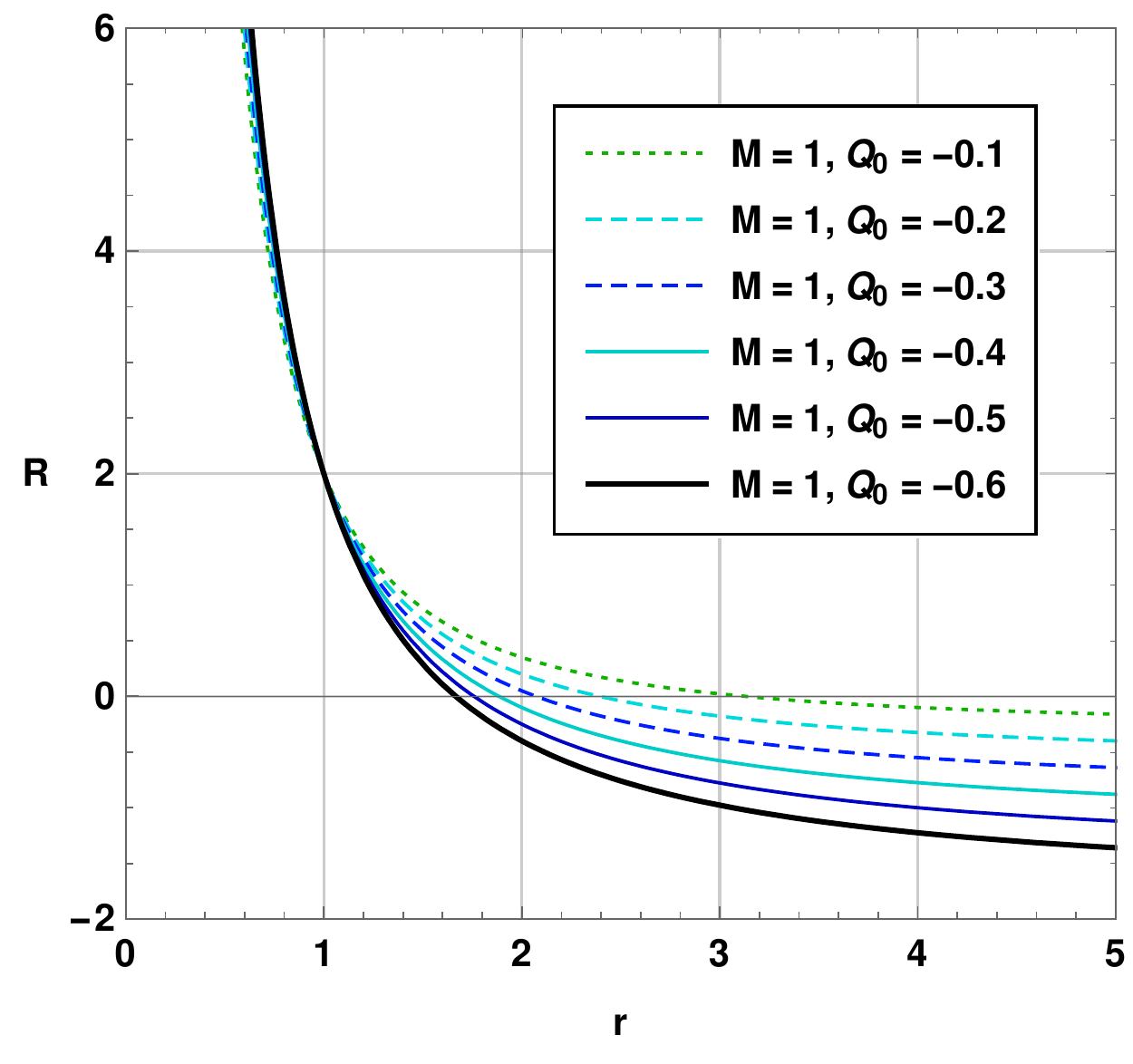}\hspace{0.5cm}
   \includegraphics[scale = 0.37]{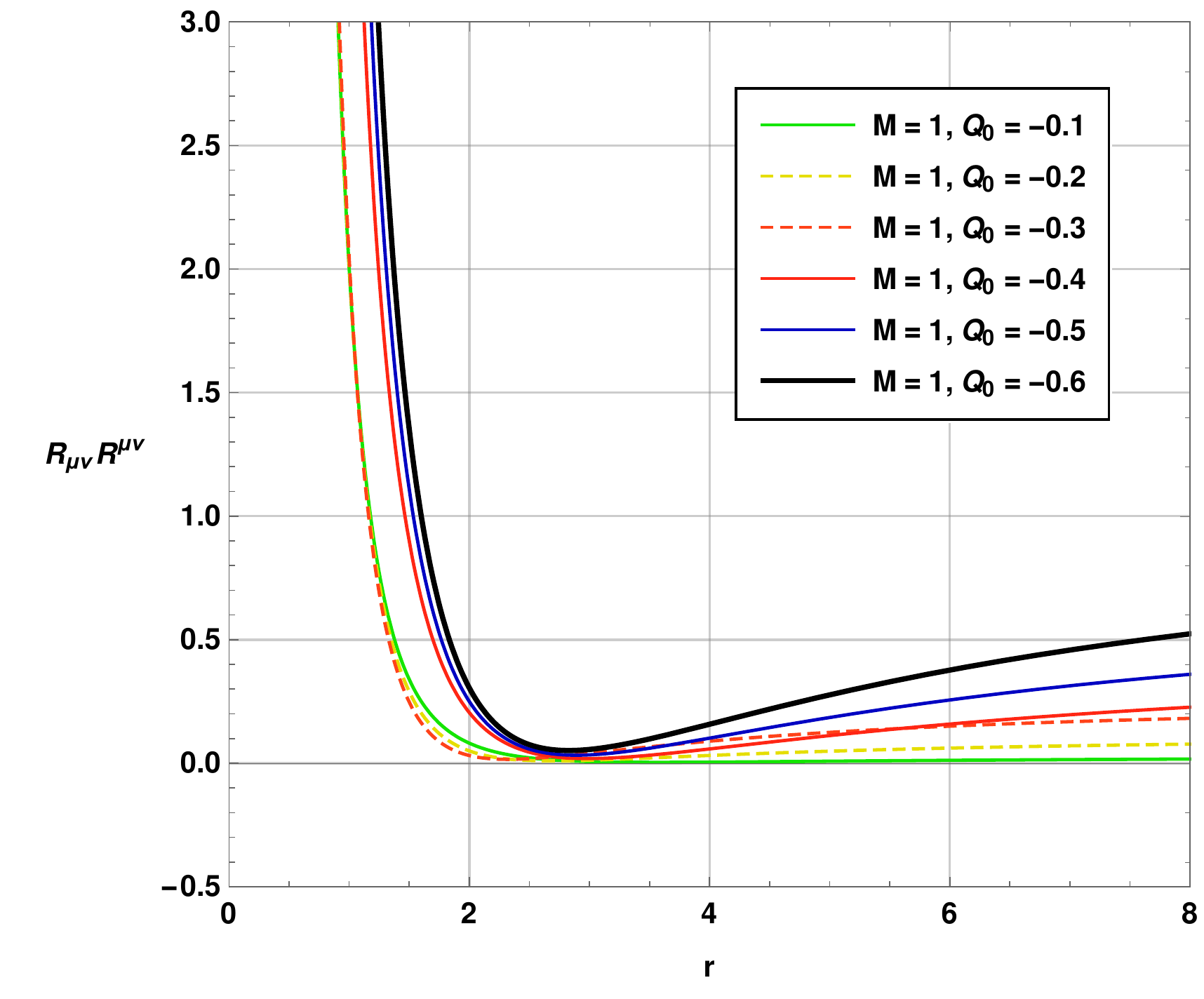}}  
\centerline{
   \includegraphics[scale = 0.35]{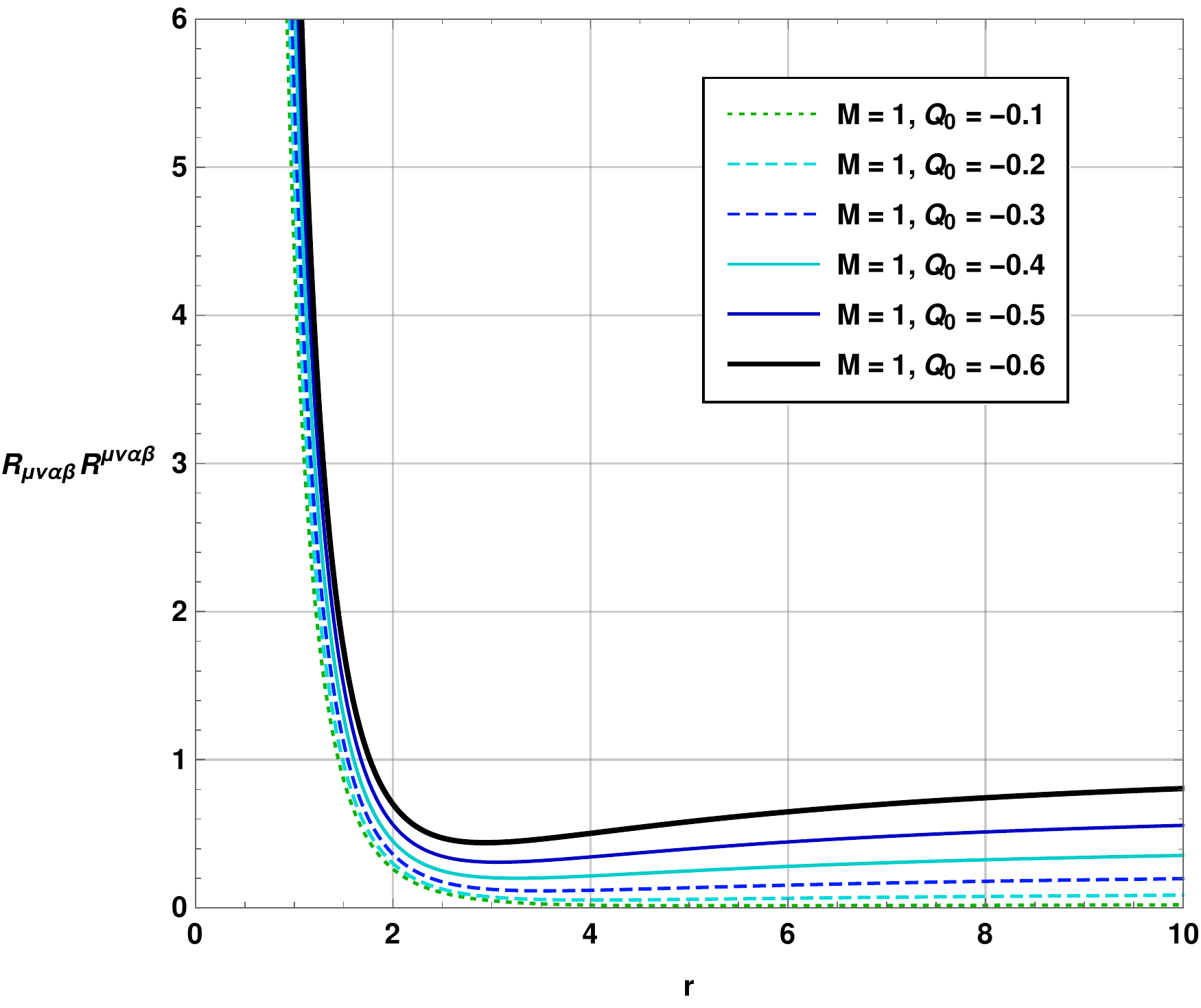}} \vspace{-0.2cm}
\caption{Variation of Ricci, Ricci squared and Kretschmann scalar w.r.t.\ $r$%
. }
\label{scalars}
\end{figure}

In this section, we perform a brief study on the black hole's
properties by studying the associated scalars. The simplest scalar
associated with the black hole spacetime is the Ricci scalar which is given
by the following expression: 
\begin{equation}
R = \frac{-3 M Q_0 r+3 Q_0 r^2+2}{r^2}.
\end{equation}
This shows that the Ricci scalar of the black hole solution depends on the
constant nonmetricity scalar $Q_0$, and for a fixed mass and at a fixed
distance, it varies linearly with the nonmetricity scalar. Another scalar
associated with the black hole spacetime is the Ricci squared scalar. The
expression for the Ricci squared scalar can be given by, 
\begin{equation}
R_{\mu\nu}R^{\mu\nu}= \frac{Q_0 r \left(9 M^2 Q_0 r-2 M \left(7 Q_0
r^2+4\right)+6 Q_0 r^3+8 r\right)+4}{2 r^4}.
\end{equation}
Unlike the Ricci scalar, Ricci squared varies nonlinearly with the
nonmetricity scalar. Finally, we calculate the Kretchmann scalar, which is
given by 
\begin{equation}
R_{\mu\nu\alpha\beta}R^{\mu\nu\alpha\beta} = \frac{Q_0 r \left(9 M^2 Q_0 r-8
M \left(Q_0 r^2+1\right)+3 Q_0 r^3+4 r\right)+4}{r^4}.
\end{equation}
One may note that this black hole has a physical singularity at $r=0$, and
the nonmetricity scalar significantly modifies the scalars associated with
the black hole spacetime. For a better visualisation of how the nonmetricity
scalar impacts these black hole space-time scalars, we have plotted them
with respect to $r$ for different values of $Q_0$. We plotted the Ricci
scalar on the first panel of Fig. \ref{scalars}. One can see that for
smaller values of the nonmetricity scalar parameter $Q_0$, the Ricci scalar
becomes negative outside the event horizon of the black hole. The scalar
vanishes at a large distance from the black hole spacetime, showing
Minkowski spacetime. On the second panel, we have considered Ricci squared
scalar $R_{\mu\nu} R^{\mu\nu}$ with respect to $r$. One can see that the
Ricci squared is positive, and when $Q_0$ decreases, it increases slowly.
Finally, we plot the Kretschmann scalar on the third panel of Fig. \ref%
{scalars}. This scalar is also positive, and with an increase in the value
of the nonmetricity scalar $Q_0$, it decreases slowly. The analysis of the
scalars shows that the nonmetricity scalar modifies this black hole
spacetime from the standard Schwarzschild black hole spacetime. The black
hole solution is unique in nature, and it has a physical singularity at $r=0$%
, which can't be avoided.

\section{Perturbations and Quasinormal modes}\label{sec5} 

We have obtained the black hole solutions in $f(Q)$
gravity. Now, in this section, we shall deal with two types of perturbations
in the black hole spacetime \textit{viz.}, massless scalar perturbation and
electromagnetic perturbation. Here we shall assume that the test field 
\textit{i.e.} scalar field or electromagnetic field has negligible impact on
the black hole spacetime. To obtain the quasinormal modes, we derive
Schr\"odinger-like wave equations for each case by considering the
corresponding conservation relations on the concerned spacetime. The
equations will be of the Klein-Gordon type for scalar fields and the Maxwell
equations for electromagnetic fields. To calculate the quasinormal modes, we
use two different methods \textit{viz.}, Bernstein spectral method and
Pad\'e averaged 6th order WKB approximation method.

Taking into account the axial perturbation only, one may write the perturbed
metric in the following way \cite{lopez2020} 
\begin{equation}  \label{pert_metric}
ds^2 = -\, |g_{tt}|\, dt^2 + r^2 \sin^2\!\theta\, (d\phi - p_1(t,r,\theta)\,
dt - p_2(t,r,\theta)\, dr - p_3(t,r,\theta)\, d\theta)^2 + g_{rr}\, dr^2 +
r^2 d\theta^2,
\end{equation}
here the parameters $p_1, p_2$ and $p_3$ define the perturbation introduced
to the black hole spacetime. The metric functions $g_{tt}$ and $g_{rr}$ are
the zeroth order terms, and they depend on $r$ only.

\subsection{Scalar Perturbation}

We start by taking into account a massless scalar field in the vicinity of
the previously established black hole. Since we assumed that the scalar
field's effect on spacetime is negligible, the perturbed metric Eq. %
\eqref{pert_metric} may be reduced to the following form: 
\begin{equation}
ds^2 = -\,|g_{tt}|\, dt^2 + g_{rr}\, dr^2 +r^2 d \Omega^2.
\end{equation}
Now, one can write the Klein-Gordon equation in curved spacetime for this
scenario as follows: 
\begin{equation}  \label{scalar_KG}
\square \Phi = \dfrac{1}{\sqrt{-g}} \partial_\mu (\sqrt{-g} g^{\mu\nu}
\partial_\nu \Phi) = 0.
\end{equation}
This equation describes the quasinormal modes associated with the scalar
perturbation. It is possible to decompose the scalar field as 
\begin{equation}
\Phi(t,r,\theta, \phi) = \dfrac{1}{r} \sum_{l,m} \psi_l(t,r) Y_{lm}(\theta,
\phi),
\end{equation}
where we have used spherical harmonics and $l$ and $m$ are the associated
indices. The function $\psi_l(t,r)$ is the radial time-dependent wave
function. One can use this equation and Eq. \eqref{scalar_KG} to have 
\begin{equation}  \label{radial_scalar}
\partial^2_{r_*} \psi(r_*)_l + \omega^2 \psi(r_*)_l = V_s(r) \psi(r_*)_l,
\end{equation}
here $r_*$ is expressed as 
\begin{equation}  \label{tortoise}
\dfrac{dr_*}{dr} = \sqrt{g_{rr}\, |g_{tt}^{-1}|}
\end{equation}
which is known as the tortoise coordinate. $V_s(r)$ stands for the effective
potential having the following explicit form: 
\begin{equation}  \label{Vs}
V_s(r) = |g_{tt}| \left( \dfrac{l(l+1)}{r^2} +\dfrac{1}{r \sqrt{|g_{tt}|
g_{rr}}} \dfrac{d}{dr}\sqrt{|g_{tt}| g_{rr}^{-1}} \right).
\end{equation}
Here, $l$ is referred to as the multipole moment of the black hole's
quasinormal modes.

\subsection{Electromagnetic Perturbation}

Now we move to the electromagnetic perturbation, where one needs to utilise
the standard tetrad formalism \cite{chandrasekhar, lopez2020} in which a
basis $e^\mu_{a}$ is defined related with the black hole metric $g_{\mu\nu}$%
. This basis satisfies, 
\begin{align}
e^{(a)}_\mu e^\mu_{(b)} &= \delta^{(a)}_{(b)}  \nonumber \\
e^{(a)}_\mu e^\nu_{(a)} &= \delta^{\nu}_{\mu}  \nonumber \\
e^{(a)}_\mu &= g_{\mu\nu} \eta^{(a)(b)} e^\nu_{(b)}  \nonumber \\
g_{\mu\nu} &= \eta_{(a)(b)}e^{(a)}_\mu e^{(b)}_\nu = e_{(a)\mu} e^{(a)}_\nu.
\end{align}
One can express tensor fields in terms of this basis as shown below: 
\begin{align*}
S_\mu &= e^{(a)}_\mu S_{(a)}, \\
S_{(a)} &= e^\mu_{(a)} S_\mu, \\
P_{\mu\nu} &= e^{(a)}_\mu e^{(b)}_\nu P_{(a)(b)}, \\
P_{(a)(b)} &= e^\mu_{(a)} e^\nu_{(b)} P_{\mu\nu}.
\end{align*}

The Bianchi identity of the field strength \textcolor{black}{$F_{[(a)(b)|(c)]} = 0$}, in the
case of the electromagnetic perturbation in the tetrad formalism results 
\begin{align}
\left( r \sqrt{|g_{tt}|}\, F_{(t)(\phi)}\right)_{,r} + r \sqrt{g_{rr}}\,
F_{(\phi)(r), t} &=0,  \label{em1} \\
\left( r \sqrt{|g_{tt}|}\, F_{(t)(\phi)}\sin\theta\right)_{,\theta} + r^2
\sin\theta\, F_{(\phi)(r), t} &=0.  \label{em2}
\end{align}
One can write the conservation equation as 
\begin{equation}
\eta^{(b)(c)}\! \left( F_{(a)(b)} \right)_{|(c)} =0,
\end{equation}
 which can be further rewritten in terms of the spherical polar coordinates as 
\begin{equation}  \label{em3}
\left( r \sqrt{|g_{tt}|}\, F_{(\phi)(r)}\right)_{,r} + \sqrt{|g_{tt}| g_{rr}}%
\, F_{(\phi)(\theta),\theta} + r \sqrt{g_{rr}}\, F_{(t)(\phi), t} = 0.
\end{equation}
 In the above expressions, a vertical rule and a comma denote the intrinsic and directional derivative with respect to the tetrad indices, respectively.
Using Eq.s \eqref{em1} and \eqref{em2}, and the time derivative of Eq. %
\eqref{em3} one gets, 
\begin{equation}  \label{em4}
\left[ \sqrt{|g_{tt}| g_{rr}^{-1}} \left( r \sqrt{|g_{tt}|}\, \mathcal{F}
\right)_{,r} \right]_{,r} + \dfrac{|g_{tt}| \sqrt{g_{rr}}}{r} \left( \dfrac{%
\mathcal{F}_{,\theta}}{\sin\theta} \right)_{,\theta}\!\! \sin\theta - r 
\sqrt{g_{rr}}\, \mathcal{F}_{,tt} = 0,
\end{equation}
where $\mathcal{F} = F_{(t)(\phi)} \sin\theta.$ Using the Fourier
decomposition $(\partial_t \rightarrow -\, i \omega)$ and field
decomposition $\mathcal{F}(r,\theta) = \mathcal{F}(r) Y_{,\theta}/\sin\theta,
$ where $Y(\theta)$ is the Gegenbauer function one can write Eq. \eqref{em4}
in the following form: 
\begin{equation}  \label{em5}
\left[ \sqrt{|g_{tt}| g_{rr}^{-1}} \left( r \sqrt{|g_{tt}|}\, \mathcal{F}
\right)_{,r} \right]_{,r} + \omega^2 r \sqrt{g_{rr}}\, \mathcal{F} -
|g_{tt}| \sqrt{g_{rr}} r^{-1} l(l+1)\, \mathcal{F} = 0.
\end{equation}
Using the definition $\psi_e \equiv r \sqrt{|g_{tt}|}\, \mathcal{F}$ and Eq. %
\eqref{tortoise}, it is possible to express Eq. \eqref{em5} in the
Schr\"odinger like form as 
\begin{equation}
\partial^2_{r_*} \psi_e + \omega^2 \psi_e = V_e(r) \psi_e,
\end{equation}
where the potential has the following explicit form: 
\begin{equation}  \label{Ve}
V_e(r) = |g_{tt}|\, \dfrac{l(l+1)}{r^2}.
\end{equation}

 One may note that due to the behaviour of electromagnetic perturbation as shown above, the potential has a simple form and a comparison with Eq. \eqref{Vs} shows that an additional term is absent in the case of the electromagnetic perturbation. This comparison allows us to combine both of them in a compact form or representation as shown below:

\begin{equation}  \label{Vcompact}
V(r) = |g_{tt}| \left( \dfrac{l(l+1)}{r^2} +\dfrac{(1-s)}{r \sqrt{|g_{tt}|
g_{rr}}} \dfrac{d}{dr}\sqrt{|g_{tt}| g_{rr}^{-1}} \right),
\end{equation}
where the new parameter $s$ stands for spin. In the case of scalar perturbation, we have spin $s=0$ which reduces Eq. \eqref{Vcompact} to Eq. \eqref{Vs}. In the case of electromagnetic perturbation, we have $s=1$, which results in the vanishing of the second term of the right-hand side of Eq. \eqref{Vcompact} and reduces it to Eq. \eqref{Ve}.

\subsection{Behaviour of Potentials}

Here we shall briefly study the behaviour of the perturbation potential for
the black hole defined above. Since the behaviour of the potential is highly
associated with the quasinormal modes, one can have a preliminary idea
related to the quasinormal modes from the potential behaviour.

One may note that the potential associated with an electromagnetic
perturbation in the normal coordinate system does not depend on the model
parameter $Q_0$, and it is completely identical to Schwarzschild's case.
However, in the case of quasinormal modes, we shall see $Q_0$ dependency due
to the tortoise coordinates, which exhibit $Q_0$ dependency. Here, we shall
only study the potential associated with scalar perturbation.

\begin{figure}[!h]
\centerline{
   \includegraphics[scale = 0.8]{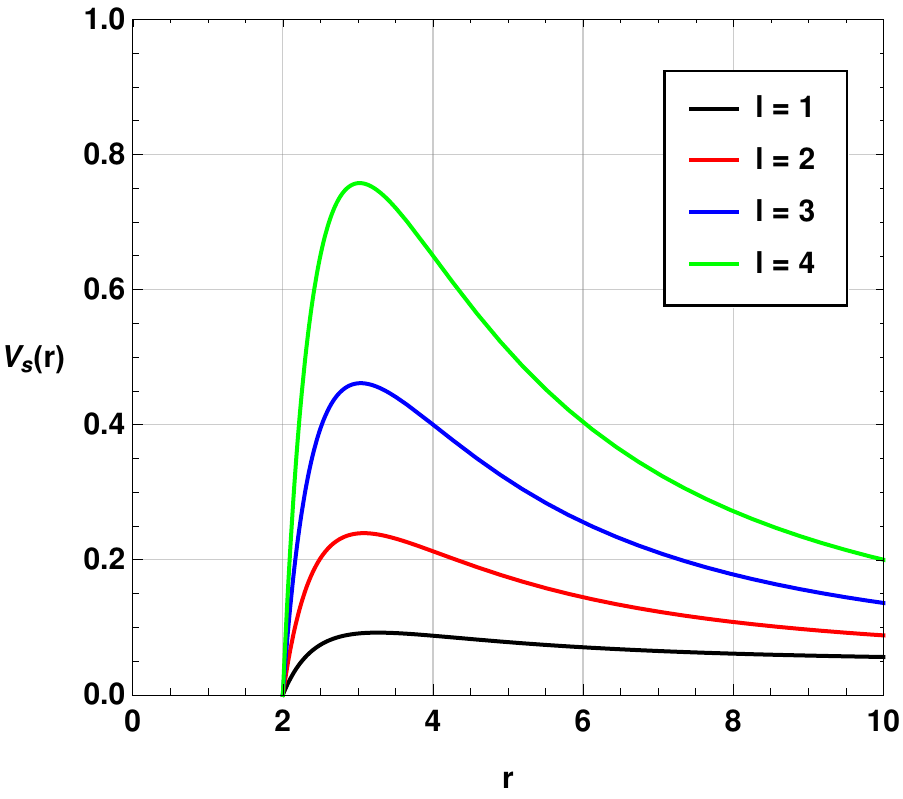}\hspace{0.5cm}
   \includegraphics[scale = 0.8]{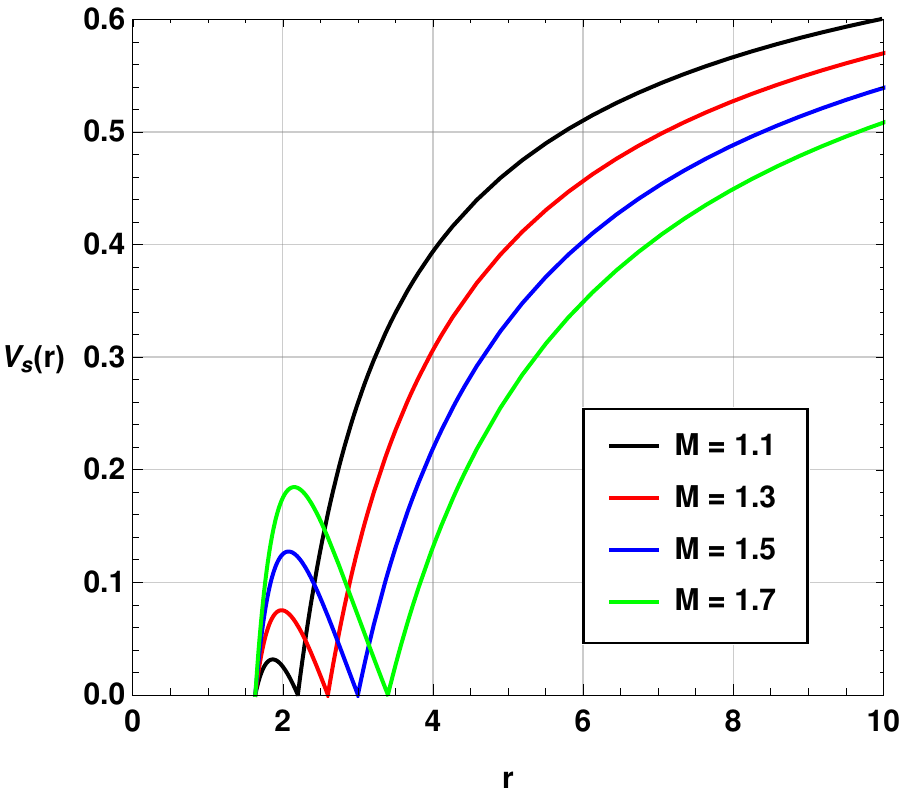}} \vspace{-0.2cm}
\caption{Variation of scalar potential $V_s(r)$ w.r.t.\ $r$. On the first panel, we have used $Q_0 = -0.1$ and $M=1$. On the second panel, we have used $Q_0 = -1.5$ and $l = 1$. }
\label{fig_Vs_01}
\end{figure}

\begin{figure}[!h]
\centerline{
   \includegraphics[scale = 0.8]{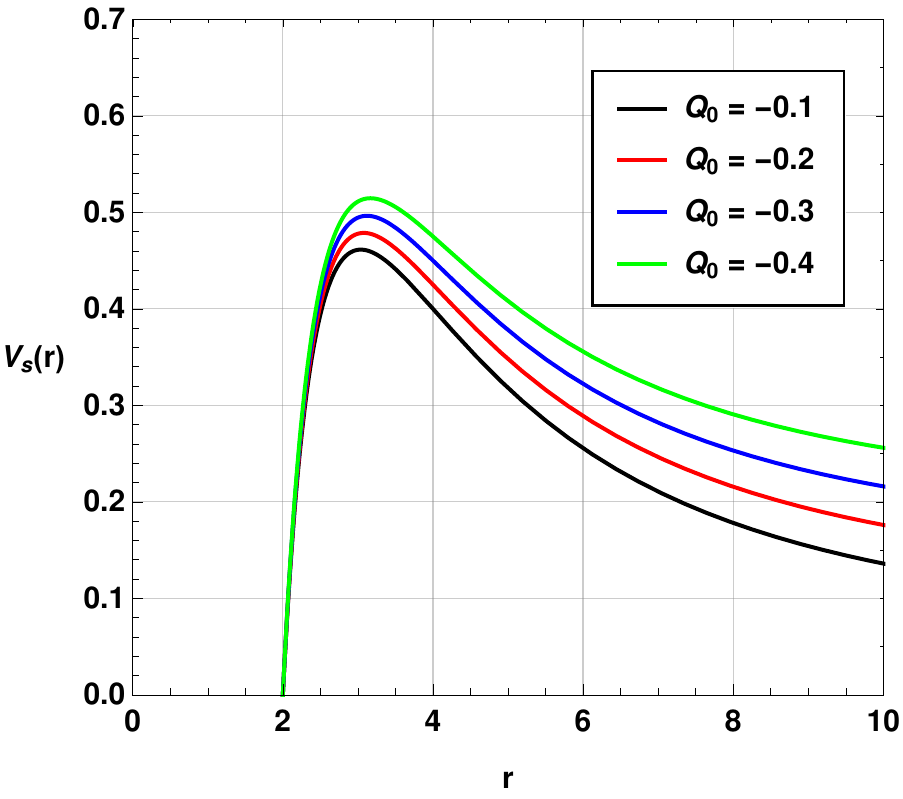}\hspace{0.5cm}
   \includegraphics[scale = 0.8]{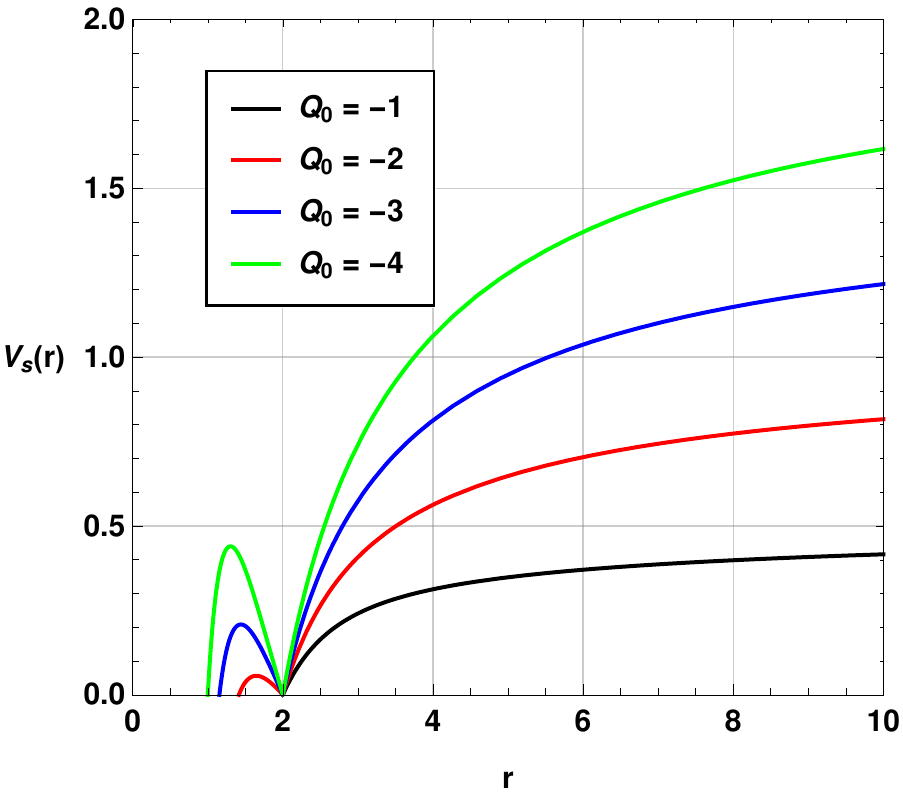}} \vspace{-0.2cm}
\caption{Variation of scalar potential $V_s(r)$ w.r.t.\ $r$. On the first panel, we have used $l = 3$ and $M=1$. On the second panel, we have used $l = 1$ and $M = 1$. }
\label{fig_Vs_02}
\end{figure}

On the first panel of Fig. \ref{fig_Vs_01}, we have shown the dependency of the scalar potential with respect to the multipole moment $l$. On the second panel, we show the impact of black hole mass on the scalar potential. One can see that the potential shows a different behaviour here which is basically due to smaller values of the parameter $Q_0$ and multipole moment $l$. We observe that the peak of the potential increases initially with an increase in the value of $M$. However, at a large distance $r$, a smaller value of $M$ corresponds to a large value of potential. To have a clear idea of how $Q_0$ impacts the scalar potential, we have plotted the potential for different values of $Q_0$ in Fig. \ref{fig_Vs_02}. On the first panel, we choose smaller values of $Q_0$ and $l=3$. On the second panel, we choose large values of $Q_0$ with $l=1$. One can see that for $l=3$, on the first panel, we observe a normal behaviour of potential. Here, with an increase in the value of $Q_0$, potential decreases. While on the second panel, we observe peaks on the potential curve for small values of $r$. With an increase in $r$, the peaks vanish and the potential decreases to $0$ at $r=2$ and beyond this point, potential increases again. In this case also, for smaller values of $Q_0$, the potential has a larger value.

This analysis shows that the model parameter $Q_0$ has a significant impact on the potential behaviour. This suggests that the parameter $Q_0$ may have noticeable impacts on the quasinormal mode spectrum.

\subsection{Bernstein Spectral method for Quasinormal modes}
The Bernstein spectral method is a numerical technique used to calculate quasinormal modes of black holes and other objects in GR and MGTs \cite{Fortuna2020}. The Bernstein method is based on a spectral decomposition of the wave equation into a finite set of basis functions, which allows for highly accurate and efficient computation of quasinormal modes \cite{Fortuna2020}.

Such a method works by approximating the wave equation as a linear combination of basis functions, such as Bernstein polynomials or Chebyshev polynomials. The coefficients of this linear combination are then calculated using a collocation method, where the wave equation is evaluated at a set of discrete points in the domain of interest. By doing this, the wave equation is transformed into a matrix eigenvalue problem, which can be solved using standard numerical techniques such as the Arnoldi method or the Lanczos algorithm. The eigenvalues obtained in this way correspond to the complex frequencies of the quasinormal modes \cite{Fortuna2020}.

One of the key advantages of the Bernstein method is that it allows for a very high degree of accuracy, since the basis functions are globally defined and can be tailored to the specific problem at hand. Additionally, the method is computationally efficient, as it only requires the solution of a matrix eigenvalue problem, which can be done very quickly using specialized algorithms. The Bernstein method has been applied to a wide range of problems in both astrophysics and mathematical physics, including the calculation of quasinormal modes for black holes of different masses and spins, the computation of wave scattering by black holes and other objects, and the study of gravitational wave generation by binary black hole systems.

In summary, the Bernstein spectral method is a powerful and versatile numerical tool for the calculation of quasinormal modes in GR as well as MGTs and other areas of physics. Its combination of accuracy and efficiency make it a valuable tool for exploring the fundamental properties of black holes and other gravitational systems, and for the study of gravitational waves and their sources.

In this subsection, we shall implement the Bernstein spectral method \cite%
{Fortuna2020} to obtain the quasinormal modes for the black hole system
considered in this study. The method has been thoroughly discussed in Ref. 
\cite{Fortuna2020}. In another recent study, this method has been revisited
for asymptotically de Sitter, anti-de Sitter and flat black hole spacetimes.
The corresponding quasinormal modes have been compared with those obtained
using time domain analysis \cite{KonoplyaB}. We define a compact coordinate $%
u$ as given by, 
\begin{equation}
u = \frac{r_h}{r}.
\end{equation}
We also define a wavefunction $\Psi(u)$ which is regular in the range $0\le
u \le 1$, as 
\begin{equation}  \label{regular_wave}
\Psi(u) = e^{i \Omega_c(\omega)/u} u^{\alpha(\omega)} (1-u)^{-i
\Omega_+(\omega)} \psi(u),
\end{equation}
where $\omega$ is the quasinormal frequency, and the other unknown
parameters can be determined from the characteristic equations \cite%
{KonoplyaB}. $\Omega_c$ and $\Omega_+$ satisfy the conditions 
\begin{equation}
\frac{d \Omega_c}{d \omega} > 0, \;\;\; \frac{d \Omega_+}{d \omega} > 0.
\end{equation}

 The quasinormal boundary conditions representing incoming and outgoing waves are
\begin{eqnarray}
\Psi\sim\left\{
  \begin{array}{rcl}
    e^{-i \omega r_\star}, \quad & r\to r_+ & (r_\star \to -\infty), \\
    e^{+i \omega r_\star}, \quad & r\to r_c & (r_\star \to +\infty).
  \end{array}
\right.
\end{eqnarray}
Here $r_\star$ implies tortoise coordinate.

In our case, we express the wavelike equation in terms of $u = 1/r$ so that $u = 0$ and $u =1$ correspond to the event horizon and cosmological horizon respectively. After that, we factorise the equation to include $e^{i \Omega_c(\omega)/u} u^{\alpha(\omega)} (1-u)^{-i
\Omega_+(\omega)}$ so that we obtain Eq. \eqref{regular_wave}. This makes the new function regular at $u=0$ and $u=1.$ This is of course due to the inclusion of $u$ and $(1-u)$ in the function. Hence, both $u$ and $(1-u)$ play a very important role. More specifically, the new function at the boundaries becomes:
$$\Psi(u=0) = 0 \; \text{and} \; \Psi(u=1) = 0,$$
which are both regular expressions providing mathematical feasibility to do the calculations numerically.
The unknown terms such as $\Omega_c(\omega)$, $\alpha(\omega)$ and $\Omega_+(\omega)$ are obtained from the characteristic equations utilising series expansion near $u=0$ and $u=1$.

The function $\psi(u)$ is represented as a sum 
\begin{equation}  \label{Bsum}
\psi(u) = \sum^N_{k=0} C_k B^N_k (u),
\end{equation}
in the above expression 
\begin{equation}
B_k^N(u)\equiv\frac{N!}{k!(N-k)!)}u^k(1-u)^{N-k}
\end{equation}
are known as the Bernstein polynomials. We use \eqref{regular_wave} in %
\eqref{radial_scalar} and then a Chebyshev collocation grid \cite{KonoplyaB}%
, 
\begin{equation}
u_p=\frac{1-\cos \frac{p\cdot\pi}{N}}{2}=\sin^2\frac{p\cdot\pi}{2N},
\end{equation}
where $p=0,...,N$, to get a number of linear equations. Now the differential
equation is simplified to the numerically solvable eigenvalue problem of a
matrix of order 2 with respect to $\omega$. The solution gives the
quasinormal frequencies, and by calculating the relevant coefficients $C_k$
and explicitly determining the polynomial \eqref{Bsum}, which roughly
approximates the solution to the wave equation, one can get the complete
solution of the problem. We compare the eigenfrequencies and related
approximating polynomials for various values of $N$ in order to rule out the
erroneous eigenvalues that emerge as a result of the polynomial basis'
finiteness. For this purpose, we follow Ref. \cite{KonoplyaB} and calculate 
\begin{equation} \label{sinfactor}
1-\frac{|\langle \phi^{(1)}\;|\;\psi^{(2)} \rangle|^2}{||\psi^{(1)}||^2||%
\psi^{(2)}||^2}=\sin^2\alpha,
\end{equation}
where $\alpha$ is the angle between two nearby eigenfunctions $\psi^{(1)}$
and $\psi^{(2)}$. We set a minimum cut-off $\alpha$ to determine the
quasinormal modes from the eigenvalues. For better accuracy, one needs to
consider a large value of $N$, which increases the computation time.

\subsection{WKB method with Pad\'e Approximation for Quasinormal modes}

Apart from the above method, we shall use another well-established method
known as the WKB method to estimate the quasinormal modes of the black hole
considered in this study. We shall compare the results obtained from both
methods to verify our findings.

The first-order WKB method or technique was introduced for the first time by
Schutz and Will in Ref.\ \cite{Schutz}. Although this method can give
approximate results of quasinormal modes, the error associated with this
method is comparatively higher. For this reason, higher-order WKB methods
have been implemented in the study of black hole quasinormal modes. The WKB
method was upgraded to higher orders in Ref.s \cite{Will_wkb, Konoplya_wkb,
Maty_wkb}. It was stated in Ref. \cite{Maty_wkb} that the WKB technique may
be improved by averaging the Pad\'e approximations. Subsequently, it was
discovered that this improved the findings of quasinormal modes with more
precision \cite{Konoplya_wkb}. The Pad\'e averaged 6th-order WKB
approximation approach will be used in this investigation.

Each table shows the quasinormal modes obtained from Pad\'e averaged $6$th
order WKB approximation method and Bernstein spectral method. The errors
associated with the WKB method are represented by the rms error $%
\vartriangle_{rms}$ and $\Delta_6$, which is defined as \cite{Konoplya_wkb} 
\begin{equation}
\Delta_6 = \dfrac{\vline \; \omega_7 - \omega_5 \; \vline}{2},
\end{equation}
where the terms $\omega_7$ and $\omega_5$ represent quasinormal modes
obtained from $7$th and $5$th order Pad\'e averaged WKB method.

\begin{table}[ht]
\caption{Quasinormal modes of the black hole with $n= 0$, $M=1$ and $Q_0=-0.1
$ for the massless scalar perturbation.}
\label{tab01}
\begin{center}
{\small 
\begin{tabular}{|cccccc|}
\hline
\;\;$l$ & \;\;Bernstein Spectral method \;\; & \;\; Pad\'e averaged WKB\;\;
& $\vartriangle_{rms}$ & $\Delta_6$ & $\Delta_{BW}$ \\ \hline
$l=1$ & $0.2929559 - 0.0601751 i$ & $0.294698\, -0.057859 i$ & $%
1.74634\times10^{-6}$ & $0.0000458873$ & $0.965011\%$ \\ 
$l=2$ & $0.4848221 - 0.0624688 i$ & $0.485050\, -0.0622951 i$ & $%
7.66312\times10^{-8}$ & $9.15631\times10^{-7}$ & $0.0585984\%$ \\ 
$l=3$ & $0.6764037 - 0.0634309 i$ & $0.676414 - 0.0634246 i$ & $%
1.61704\times10^{-8}$ & $7.52219\times10^{-8}$ & $0.00178328\%$ \\ 
$l=4$ & $0.8682419 - 0.0638755i$ & $0.868242 - 0.0638755 i$ & $%
4.59096\times10^{-9}$ & $1.99499\times10^{-8}$ & $0.0000082445\%$ \\ \hline
\end{tabular}
}
\end{center}
\end{table}

\begin{table}[ht]
\caption{Quasinormal modes of the black hole with $n= 0$, $M=1$ and $Q_0=-0.1
$ for the electromagnetic perturbation.}
\label{tab02}
\begin{center}
{\small 
\begin{tabular}{|cccccc|}
\hline
\;\;$l$ & \;\;Bernstein Spectral method \;\; & \;\; Pad\'e averaged WKB\;\;
& $\vartriangle_{rms}$ & $\Delta_6$ & $\Delta_{BW}$ \\ \hline
$l=1$ & $0.2655401 - \ 0.0652171i$ & $0.265539 - 0.065161 i$ & $0.0000177085$
& $6.62115\times10^{-6}$ & $0.0205383\%$ \\ 
$l=2$ & $0.4676216 - 0.0647356i$ & $0.467621 - 0.0647357 i$ & $%
7.32702\times10^{-7}$ & $4.67206\times10^{-7}$ & $0.000122958\%$ \\ 
$l=3$ & $0.6639979 - 0.0646406i$ & $0.663998 - 0.0646407 i$ & $%
9.4691\times10^{-8}$ & $7.50012\times10^{-8}$ & $0.0000147967\%$ \\ 
$l=4$ & $0.8585978 - 0.0646038i$ & $0.858598 - 0.0646038 i$ & $%
2.07449\times10^{-8}$ & $1.99993\times10^{-8}$ & $0.0000209396\%$ \\ \hline
\end{tabular}
}
\end{center}
\end{table}

In Table \ref{tab01}, we have listed the quasinormal modes for the massless
scalar perturbation with the model parameter $Q_0=-0.1$, the mass of the
black hole $M=1$ and overtone number $n=0$. The term $\Delta_{BW}$
represents the percentage deviation of the quasinormal modes obtained via
the Bernstein spectral method from the 6th order pad\'e averaged WKB
approximation method. 
 Here in the Bernstein spectral method, a large value of $N$ increases the accuracy \cite{Burikham:2017gdm} but it needs a larger computation time. We have used $N=20, 30, 40$ and $60$ to compare the results for excluding the spurious eigenvalues. We compare the results by extracting the eigenvalues from each set that differ by an amount smaller than the specified cutoff value (around $0.01$). For each pair of eigenvalues meeting this criterion, we calculate the squared $\sin$ of the angle between their respective eigenfunctions using Eq. \ref{sinfactor}. When the $\sin$ value approaches zero or less than a cutoff value (we took it to be $1 \times 10^{-10}$), it signifies that the eigenfunctions exhibit a negligible difference and approximate a common solution, possibly differing by a constant factor. By employing this technique, we effectively identify the dominant quasinormal modes. We see that excluding spurious eigenvalues and considering a suitable precision of numerical calculations allow us to choose small values of $N$ say around $40$ to obtain the desired accuracy instead of considering a large value such as $N=100$ which needs a higher computation time. 
One can see that for $l=1$, $\Delta_{BW}$ has a
comparatively higher value. Apart from this, the errors associated with the
Pad\'e averaged WKB method are also higher. The percentage deviation, as
well as the errors, decrease significantly as the multipole moment $l$
increases. The variation of these errors can be associated with the WKB
method, which gives less significant results when $l-n$ is smaller. For both
methods, it is clearly seen that the quasinormal frequencies and the damping
rate increase with the value of the multipole moment $l$.

In Table \ref{tab02}, we have shown the quasinormal modes for
electromagnetic perturbation for different $l$ values with $M=1$, $n=0$, and 
$Q_0=-0.1$. One can see that in this case also, the deviation term $%
\Delta_{BW}$ has higher values for smaller multipole moment $l$. As $l$
increases, the deviation, as well as the error terms, decreases noticeably.

From both Tables, it is clear that quasinormal frequencies have lower values
for the electromagnetic perturbation than those obtained in the case of
massless scalar perturbation. However, the decay rate or damping rate of GW
is higher in the case of electromagnetic perturbation.

\begin{figure}[htbp]
\centerline{
   \includegraphics[scale = 0.8]{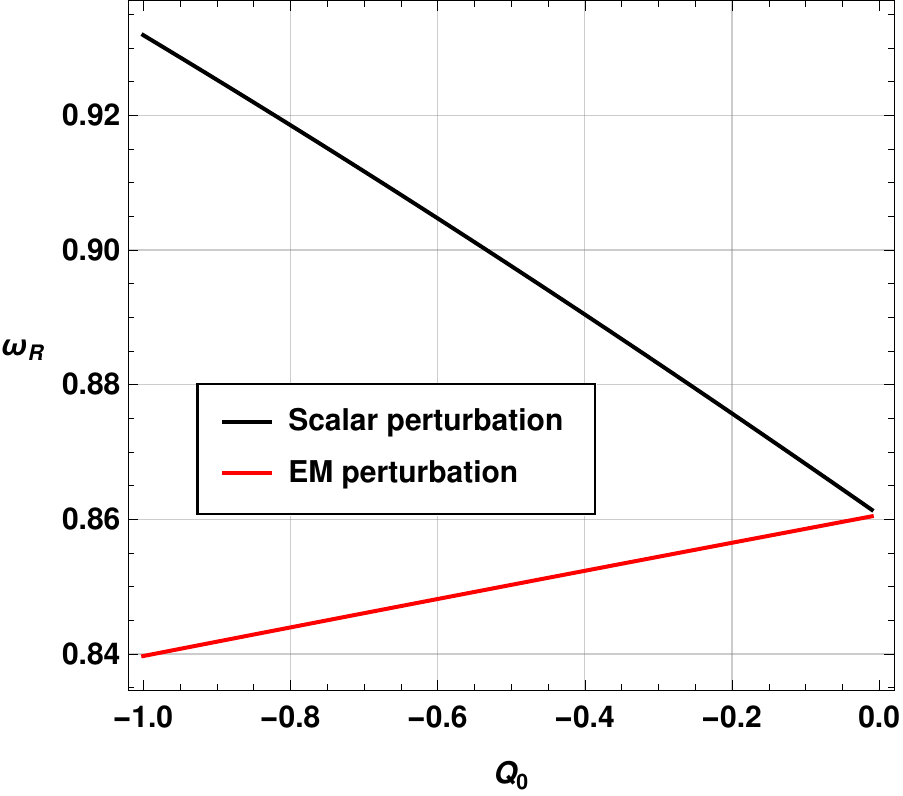}\hspace{0.5cm}
   \includegraphics[scale = 0.8]{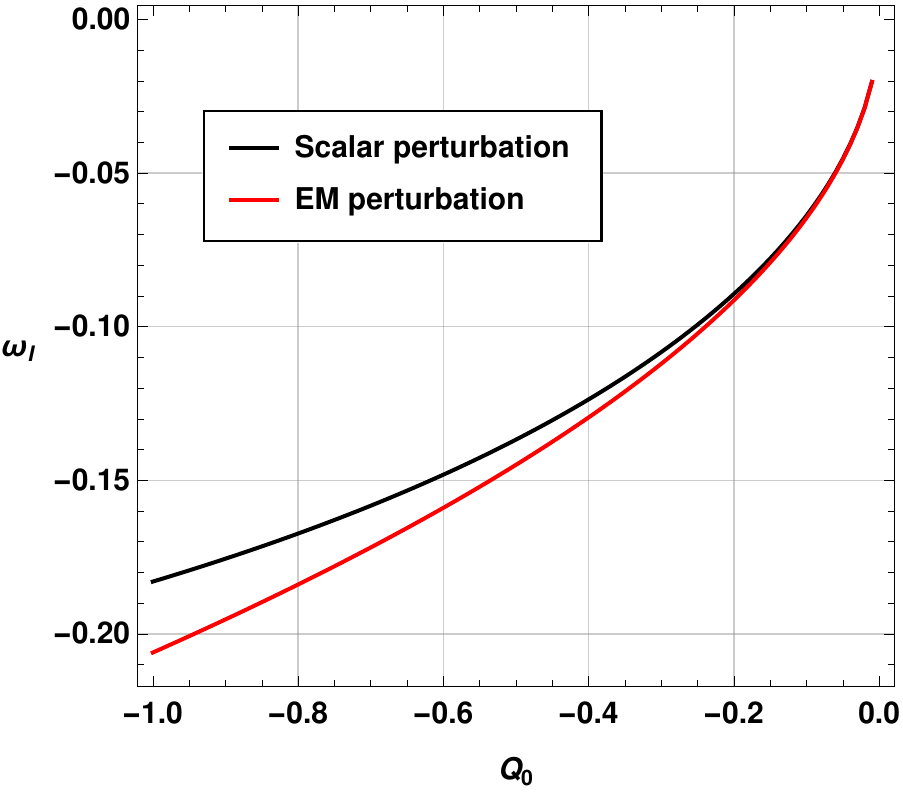}} \vspace{-0.2cm}
\caption{Variation of quasinormal modes with respect to model parameter $Q_0$
with $M=1$, $n= 0$ and $l=4$ for massless scalar perturbation and electromagnetic
perturbation.}
\label{QNMs01}
\end{figure}

To see the effect of the model parameter $Q_0$ on the quasinormal modes, we
have plotted the variations of quasinormal modes with respect to $Q_0$. On
the left panel of Fig.\ref{QNMs01}, we have plotted the real frequencies of
quasinormal modes with respect to the nonmetricity scalar $Q_0$ for both
massless scalar perturbation and electromagnetic perturbation. One can see
that the scalar quasinormal mode frequency decreases significantly with an
increase in the nonmetricity scalar $Q_0$. On the other hand,
electromagnetic quasinormal frequencies increase slowly with an increase in
the value of $Q_0$. Both types of frequencies approach each other when $Q_0$
is close to zero. One may note that $Q_0$ has more impact on the scalar
quasinormal frequencies than the electromagnetic quasinormal frequencies. On
the second panel of Fig.\ref{QNMs01}, we have plotted the imaginary
quasinormal modes with respect to the nonmetricity scalar $Q_0$. We observe
that the damping rate or decay rate of gravitational waves increases with a
decrease in the value of the scalar $Q_0$. The decay rate for the scalar
perturbation is lower than that of electromagnetic perturbation for smaller
values of $Q_0$ near $-1$. However, when $Q_0$ increases to $0$, the damping
rate decreases and approaches zero. One may note that the variation of the
damping rate with respect to $Q_0$ is nonlinear, unlike the case for real
quasinormal modes, where the variation was linear for both types of
perturbations.

Our analysis shows that the quasinormal mode spectrum carries the signature
of nonmetricity, and in the near future, provided sufficient significant
observational data is available, it might be possible to distinguish the
scalar and electromagnetic quasinormal modes as well as the presence of
nonmetricity in theory from a standard Schwarzschild black hole in GR.

\section{Evolution of Scalar and Electromagnetic Perturbations on the Black
hole geometry}\label{sec6} 

In the previous section, we have numerically calculated the quasinormal modes and studied their behaviour with respect to the nonmetricity scalar $Q_0$. In this section, we shall deal with the time domain profiles of the scalar perturbation and electromagnetic perturbation. To obtain the time evolution profiles, we shall implement the time domain integration formalism \cite{gundlach}. For this purpose, we define $\psi(r_*,
t) = \psi(i \Delta r_*, j \Delta t) = \psi_{i,j} $ and $V(r(r_*)) = V(r_*,t) =
V_{i,j}$. Now, one can express Eq. \eqref{scalar_KG} as
\begin{equation}
\dfrac{\psi_{i+1,j} - 2\psi_{i,j} + \psi_{i-1,j}}{\Delta r_*^2} - \dfrac{%
\psi_{i,j+1} - 2\psi_{i,j} + \psi_{i,j-1}}{\Delta t^2} - V_i\psi_{i,j} = 0.
\end{equation}
We set the initial conditions $\psi(r_*,t) = \exp \left[ -\dfrac{(r_*-k_1)^2%
}{2\sigma^2} \right]$ and $\psi(r_*,t)\vert_{t<0} = 0$ (here $k_1$ and $%
\sigma$ are median and width of the initial wave-packet) and then calculate the time evolution of
the scalar field as 
\begin{equation}
\psi_{i,j+1} = -\,\psi_{i, j-1} + \left( \dfrac{\Delta t}{\Delta r_*}
\right)^2 (\psi_{i+1, j + \psi_{i-1, j}}) + \left( 2-2\left( \dfrac{\Delta t%
}{\Delta r_*} \right)^2 - V_i \Delta t^2 \right) \psi_{i,j}.
\end{equation}
Using the above iteration scheme and choosing a fixed value of $\frac{\Delta t}{\Delta r_*}$, one can easily obtain the profile of $\psi$ with respect to time $t$. However, one should keep $\frac{\Delta t}{\Delta r_*}
< 1$ so that the Von Neumann stability condition is satisfied during the numerical procedure.

\begin{figure}[htbp]
\centerline{
   \includegraphics[scale = 0.8]{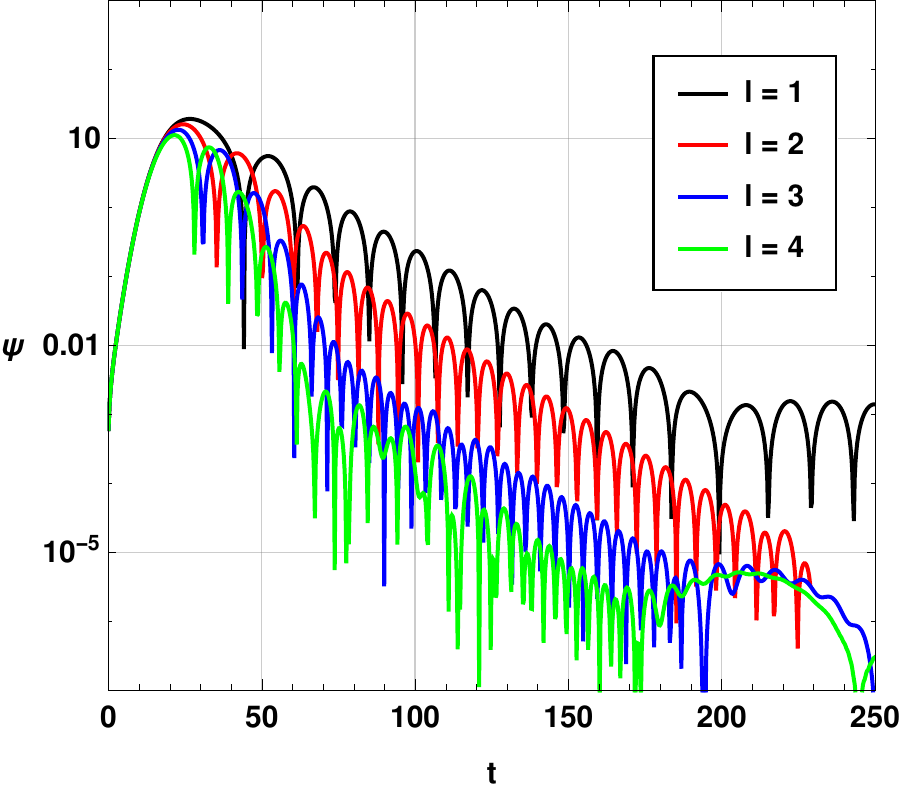}\hspace{0.5cm}
   \includegraphics[scale = 0.8]{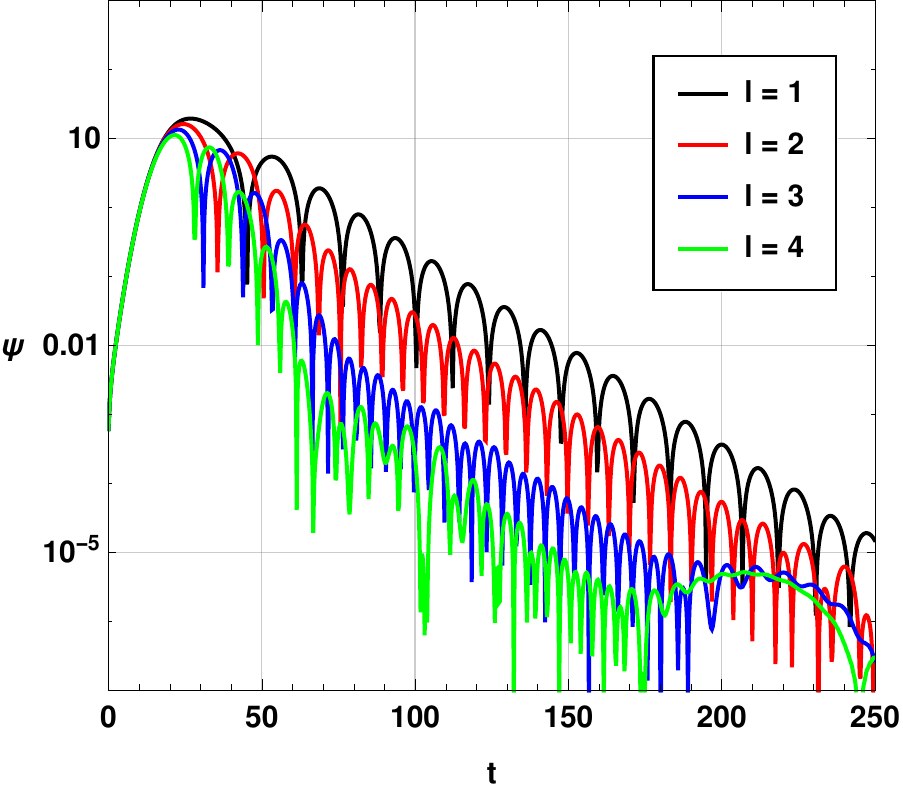}} \vspace{-0.2cm}
\caption{Time domain profiles with $M=1$, $n= 0, Q_0=-0.1, $ for massless scalar
perturbation (on first panel) and electromagnetic perturbation (on second
panel). }
\label{time01}
\end{figure}

\begin{figure}[htbp]
\centerline{
   \includegraphics[scale = 0.8]{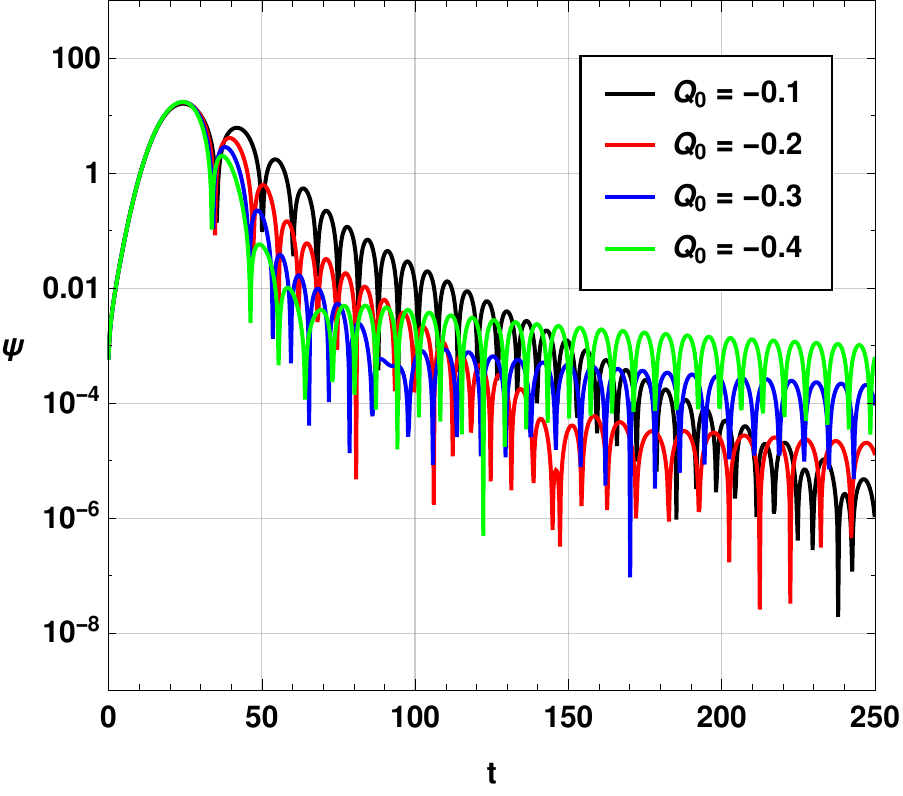}\hspace{0.5cm}
   \includegraphics[scale = 0.8]{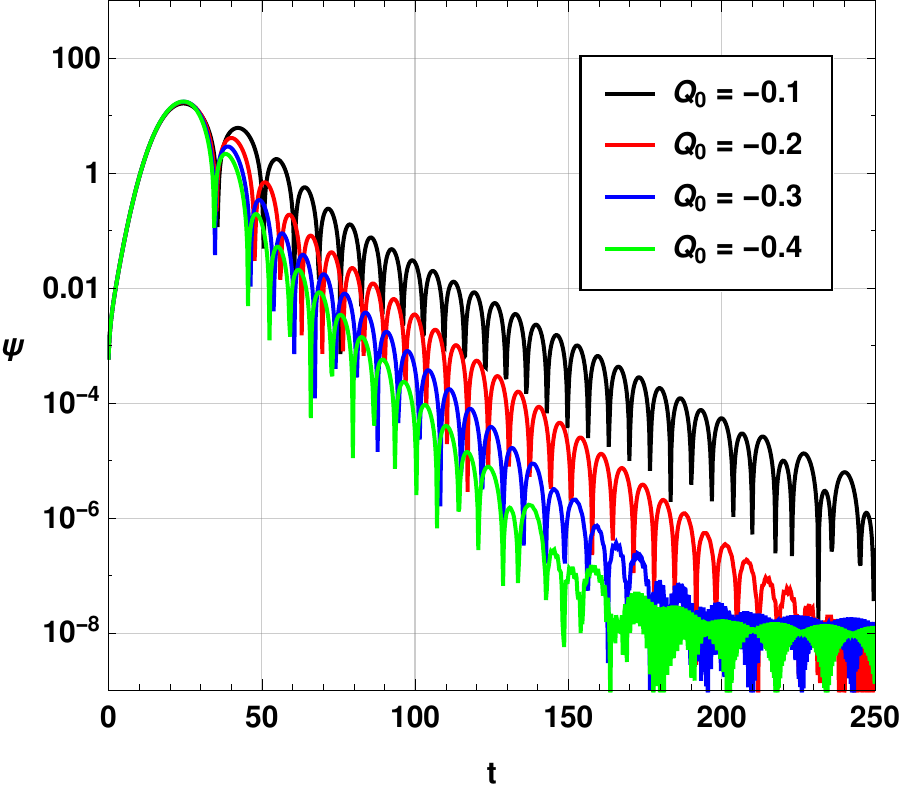}} \vspace{-0.2cm}
\caption{Time domain profiles with $M=1$, $n=0, l=2$ for massless scalar
perturbation (on first panel) and electromagnetic perturbation (on second
panel). }
\label{time02}
\end{figure}

On the left panel of Fig.\ \ref{time01}, we have shown the time domain 
profile for the scalar perturbation and on the right panel, we have shown the 
time domain profile for the electromagnetic perturbation with overtone number 
$n=0$, nonmetricity scalar constant $Q_0=-0.1$ and multipole number $l=1,2,3$ 
and $4$. We can see that the time domain profiles are significantly different 
for different values of multipole number $l$. For both cases, with an 
increase in the value of $l$, we observe an increase in the frequency. 
However, the decay rate of the time domain profile seems to increase for the 
scalar perturbation with an increase in the value of $l$ while for 
electromagnetic perturbation the variation is very small. Another interesting 
observation is that for $l=1$, both the scalar and electromagnetic time 
domain profiles have significantly different damping rate. The damping or 
decay rate seems to higher for the electromagnetic perturbation case than 
that of the scalar perturbation. 

In Fig. \ref{time02}, we have plotted the time domain profiles for scalar 
perturbation (on the left panel) and electromagnetic perturbation (on the 
right panel) with different values of the parameter $Q_0$ with overtone $n=0$ 
and multipole number $l=2$. One may note that the decay rate for the time 
profile of electromagnetic perturbation increases more rapidly with a 
decrease in the value of the parameter $Q_0$. The results obtained from the 
time domain profiles are consistent with the previous results of quasinormal 
modes.

\section{Conclusion}\label{sec7} 
In this work, we have studied the quasinormal modes of a static black hole in 
$f(Q)$ gravity. The black hole solution is obtained for the model 
$f(Q)=\underset{n}{\sum}a_{n}\left(Q-Q_{0}\right)^{n} $. We have considered 
two types of perturbations, {\it viz.,} scalar perturbation and 
electromagnetic perturbation and obtained the corresponding potentials. We 
have seen that the potential in normal coordinates for the electromagnetic 
perturbation is identical to the potential corresponding to a Schwarzschild 
black hole. However, in the tortoise coordinate, the potential depends on the 
model parameter $Q_0$. As a result, both types of perturbations depend on the 
model parameter $Q_0$. To obtain quasinormal modes, we implement a newly 
introduced method known as Bernstein spectral method. To confirm the results 
obtained from this new method, we have considered well-known Pad\'e averaged 
6th-order WKB approximation method and found that the results obtained from 
the previous method stand in agreement with the results obtained from 
the higher-order WKB method.

The quasinormal modes vary significantly with the model parameter $Q_0$. It 
is found that for a nonzero $Q_0$, real quasinormal modes for scalar 
perturbation are always higher than those for electromagnetic perturbation. 
On the other hand, for smaller values of $Q_0$, the damping rate of 
gravitational waves associated with the scalar perturbation is always lower 
than that associated with the electromagnetic perturbation. We have also 
investigated the time domain profiles of the perturbations for different 
values of the model parameters. The results obtained from the time domain 
analysis agree well with the results provided by the numerical methods.

Since the black hole solutions are different in the $f(Q)$ nonmetricity 
theory from the solutions in other theories, the properties of black holes 
may be different in the framework of such theories. If any significant 
impacts are present in the generation and propagation of gravitational waves, 
recent detection of gravitational wave events can play a significant role in 
the study of the viability of such theories. Quasinormal modes are one of the 
interesting and widely studied properties of a perturbed black hole spacetime,
and hence a detailed study of quasinormal modes from black holes in the 
framework of $f(Q)$ is necessary to understand the theory better. Our study 
shows that the black hole solution considered in the study can generate 
significantly different quasinormal modes from that of a Schwarzschild black 
hole \cite{Chandrasekhar_qnms}. Moreover, the quasinormal modes for the 
electromagnetic perturbation, as well as the scalar perturbation, are 
noticeably different for nonzero values of the model parameter $Q_0$.

The study of black holes and their properties in $f(Q)$ gravity is not very 
old, and still, there are very few studies incorporating various aspects of 
black holes in this theory. The study of different possible black hole 
solutions, especially in the presence of different promising surrounding 
relics, is necessary to understand the theory properly. Moreover, the 
thermodynamical aspects as well as quasinormal modes of black holes in $f(Q)$ 
are still left to be explored. Gravitational perturbation and related optical 
properties of the black hole, such as shadow, sparsity etc., are some probable 
future prospects of the study.

\section{ACKNOWLEDGMENTS}

A. {\"O}. would like to acknowledge the contribution of the COST Action
CA18108 - Quantum gravity phenomenology in the multi-messenger approach
(QG-MM). A. {\"O}. would like to acknowledge networking support by the COST Action CA21106 - COSMIC WISPers in the Dark Universe: Theory, astrophysics and experiments (CosmicWISPers). DJG would like to thank Prof. U. D. Goswami for some useful 
discussions.


\end{document}